\documentclass[12pt]{article}
\usepackage{graphicx,psfrag,amsmath,amsthm,amssymb,euscript}
\newcounter{aaa}
\newenvironment{teor}[2][{}]{\begin{trivlist}\refstepcounter{aaa}%
\labelsep=0pt\item[\bfseries \theaaa. #2. ]#1}%
{\end{trivlist}}
\newcommand{\ssy}[7][]{#2,  \emph{#3, #4} {\bf #5} (#6) #7 \texttt{\small #1}\rlap{.}}
\newcommand{\rmd}{\mathrm{d}}
\newcommand{\shir}{ w}
\DeclareMathAlphabet{\eu}{U}{eur}{m}{n}
\newcommand*{\s}{\eu{S}}
\DeclareSymbolFont{iso}{U}{txmia}{m}{it}
\DeclareMathSymbol{\eucl}{\mathalpha}{iso}{"85}
\DeclareMathSymbol{\cel}{\mathalpha}{iso}{"9A}
\DeclareMathSymbol{\rea}{\mathalpha}{iso}{"92}
\DeclareMathSymbol{\eucl}{\mathalpha}{iso}{"85}
\DeclareMathSymbol{\Mf}{\mathalpha}{iso}{"8D}
\newcommand{\bole}{\text{\textbf{\textit e}}}
\newcommand{\kartil}[4][]{\begin{figure}#2\begin{center}
\includegraphics[width=\textwidth]{#3}%
\end{center}#1\caption{#4}\end{figure}}

\title{Quasiregular singularities taken seriously}

\author{Serguei Krasnikov\\
        The Central Astronomical Observatory of   RAS at Pulkovo}

\date{}
\begin{document}
\maketitle
\begin{abstract}
I discuss a special class of singularities obtained as a natural
4-di\-men\-sional generalization of the conical singularity. Such
singularities (called quasiregular) are ruinous for the predictive force
of general relativity, so one often assumes (implicitly as a rule) that
they can be somehow excluded from the theory. In fact, however, attempts
to do so (without forbidding the singularities by fiat) have failed so
far. It is advisable therefore to explore the possibility that their
existence is not prohibited after all. I argue that quasiregular
singularities, if allowed, may appear either in situations where
causality is endangered or in the early Universe. In the latter case
objects might appear strongly (though not quite) resembling cosmic
strings. Those objects would be observable and, moreover, it is not
impossible that we already do observe one.
\end{abstract}
\newpage

\section{Introduction}
This lecture is devoted to a special kind of singularity. Instead of
giving a precise definition let me start with a simple example.
\psfrag{h}{$\zeta$}\psfrag{0}{0}
\psfrag{i}{identify}
\kartil[\hspace*{5em} (a)\hfill (b)\hspace*{5em}]{[h]%
\psfrag{z}{$\zeta=z,t$}\psfrag{M}{$\EuScript M$}
\psfrag{g}{$\gamma^*$}}{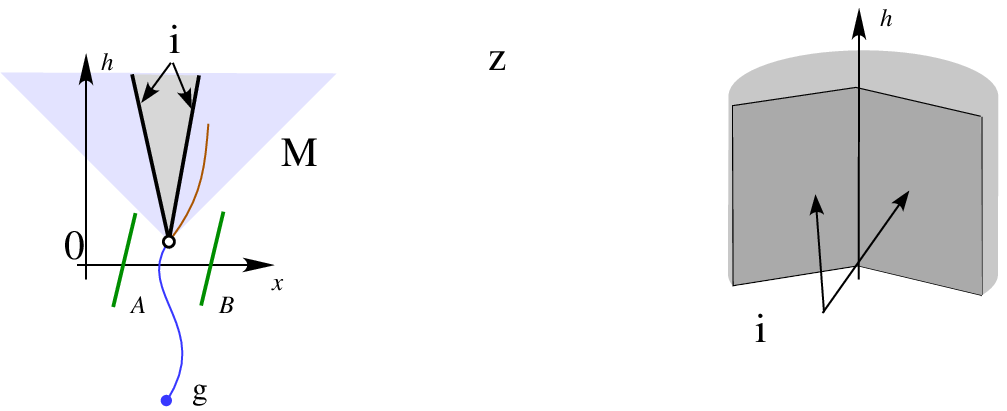}{$\zeta$
in this and subsequent figures may be
\label{fig:primerb} $z$ or $t$. a) In the Minkowski case both sides of the angle should be
timelike. b) Initially the space is $\rea^3$ with the flat metric.} Take
a two-dimensional space, either Euclidean, or Minkowskian. Cut an angle
out of it and glue the rays bounding the angle together (the vertex is
not regarded a part of a ray). The resulting cone $\EuScript M$, see
Fig.~\ref{fig:primerb}a,  is still a smooth connected paracompact flat
manifold (so, in particular, in the Lorentzian case it is a nice
spacetime)\footnote{A proof of this, hopefully obvious, assertion
requires giving a precise meaning to the term ``gluing'', see, e.~g.,
\cite{strstr}.}, but it is
\emph{singular}
--- the vertex cannot be glued back into the space (without sacrificing
either the smoothness or the non-degeneracy of the metric), so the
geodesics terminating at the hole are endless though incomplete. In
Fig.~\ref{fig:primerb}b a similar construction is shown in the two plus
one case. We again remove a wedge and glue together its boundaries (this
time they are half-planes). Again the intersection of the faces --- this
time it is the straight line
--- cannot be glued back into the space and we again have a singularity,
this time ``in the form'' of a straight line.

It is the  singularities of this type that are our subject. The reason
why they deserve  most serious consideration is that they, in fact,
deprive general relativity of its predictive power. Indeed, in contrast
to the ``usual'', curvature singularities, these ``topological'' ones are
absolutely ``sudden'':
\emph{nothing} would tell an observer approaching such a singularity that
his world line will terminate in a moment. As we shall see soon, in
spacetimes with such singularities everything (the geometry of the
universe, its topology, causal structure, etc.) may change whimsically
and (apparently) causelessly. For example, observers (like $A$ and $B$ in
the figure) may think --- up to some moment --- that they live in the
Minkowski space. But after that moment they
 will discover that without
experiencing any acceleration they acquired some speed towards each
other. Likewise, in the otherwise Minkowskian space  time machines or
wormholes may appear with no visible cause if their appearance is
accompanied by singularities of this kind. As Geroch
\cite{GerHole} stated  in this
connection: ``Thus general relativity, which seemed at first as though it
would admit a natural and powerful statement at prediction, apparently
does not''. Two hard questions that immediately arise are:
\begin{enumerate}
  \item How to predict the evolution of the Universe? We see that
  anything can happen any time.
  \item Why don't we encounter  that  problem in our everyday life?
\end{enumerate}
A temptingly simple answer would be this: ``Such singularities are
unphysical. They are excluded by\dots'' In place of the dots a strong
argument must stand, or a new (physically motivated) postulate.

\section{Inevitability of quasiregular singularities}
In this section my goal is to explain why, contrary to what one might
expect, it is hard (if possible at all) to find that appropriate argument
or postulate.
\subsection{Identification}
Before discussing possible candidates we, of course, have to specify
clearly what are ``such singularities''. This  can be done in many ways,
but I shall restrict myself to three most known variants. References to
some others can be found in \cite{hole}.
\subsubsection{Quasiregular singularities} Consider a curve in a
spacetime $M$
\[\gamma(s)\colon\quad [0,1)
\to M\]
and let $\{\bole_{(i)}\}$ be an orthonormal frame parallel transported
along $\gamma$. In this frame we find the components of $\gamma$'s
velocity  and define  the following integral:
\[
b[\gamma]\equiv \int_{0}^1
\sqrt{(\,\,\dot{\!\!\gamma}\,{}^1)^2 + (\,\,\dot{\!\!\gamma}\,{}^2)^2 +
\dots}\,\rmd\xi.
\]
called the \emph{$b$-length} of $\gamma$. In the Riemannian case it is
simply the length of the curve. On the other hand, in the Minkowski space
the $b$-length of  a curve is merely its coordinate length in the
standard (Cartesian) coordinates.

Clearly, the \emph{value} of $b$ may depend on the choice of
$\{\bole_{(i)}\}$, but its
\emph{finiteness} does not.
\begin{teor}{Definition} An inextendible spacetime is said to be singular
($b$-incomplete) if there is an \emph{endless} curve  $\gamma^*$ with a
finite $b$-length.
\end{teor}
Obviously, our exemplary spacetime $\EuScript M$ is singular: the blue
curve is endless (to be more precise, it is \emph{future} endless) even
though it has a finite $b$-length.
\begin{teor}{Definition} A singularity is \emph{quasiregular} if in the
basis $\{\bole_{(i)}\}$ the components of the Riemann tensor and all its
derivatives remain bounded on $\gamma^*$.
\end{teor}
Evidently,  singularities in flat spacetimes (including $\EuScript M$)
are quasiregular. A
\emph{general} quasiregular singularity, however, is a much more complex
object than simply a punctured plane.

\begin{teor}{Example. (Misner space)}
Take the Minkowski plane
\begin{equation}\label{eq:MinkPl}
\rmd s^2=-\rmd t^2+ \rmd x^2= -\rmd\alpha\rmd\beta
\qquad \alpha\equiv t-x ,\quad  \beta\equiv t+x ,
\end{equation}
and consider the isometry $\eta$ (it is a Lorentzian boost, in fact)
\[
\eta\colon\quad \alpha\mapsto\kappa\alpha,\quad
\beta\mapsto\kappa^{-1}\beta
 \]
This isometry induces an equivalence relation: a point $p$ is equivalent
to any point $q$ related to it by the isometry:
\[
p\approx q \quad\Longleftrightarrow \quad  p=
\eta^k(q) \qquad\forall\, k\in\cel
\]
The Misner space $M_\text{M}$ is defined, see, e.g., \cite{HawEl}, as the
quotient of the left
\emph{half-plane} $H$ over this equivalence
\psfrag{H}{$H$}
\psfrag{t}{$t$}
\psfrag{a}{$\alpha$}%
\psfrag{b}{$\beta$}%
\psfrag{a0}{$\alpha_0$}%
\psfrag{ka0}{$\kappa\alpha_0$}%
\psfrag{k2a0}{$\kappa^2\alpha_0$}%
\psfrag{k2}{$\kappa^2$}%
\psfrag{b0}{$1$}%
\psfrag{kb0}{$\kappa$}
\psfrag{hor}{$\alpha=0$}
\psfrag{g}{$\gamma$}
\psfrag{l}{$\lambda$}
\[
M_\text{M}= H/\approx.
\]
It follows right from the definition (I drop the proof) that the Misner
space is a nice legitimate spacetime, though  with a lot of surprising
properties.
\kartil[\hspace*{4em} (a)\hfill (b)\hspace*{4em}]{[h]%
\psfrag{H}{$H$}
\psfrag{t}{$t$}
\psfrag{m}{$\mu$}
}{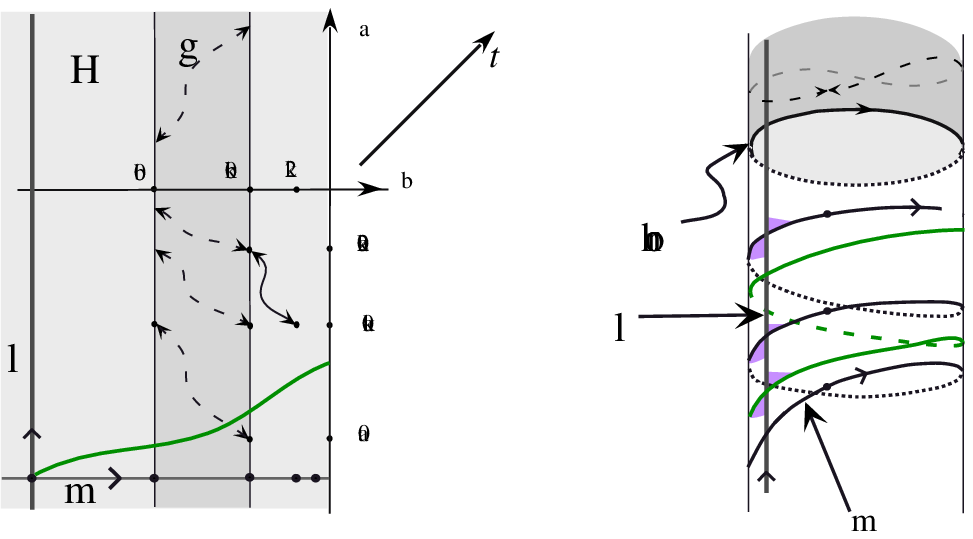}{\label{fig:misnb2} (a) The double-arrowed lines connect
(some of) identified points. (b) By purple some of the null cones are
shown.}
 To see some of them it is instructive to
isolate the strip  $\beta\in(1 ,\kappa)$ as a fundamental region. Then
the Misner space is   obtained simply by identifying the left border of
the strip with its right border according to the rule
$$(\alpha,1)= (\kappa^{-1}\alpha,\kappa) $$
as shown by dashed lines in Fig.~\ref{fig:misnb2}a. So, $M_\text{M}$ is a
cylinder. The metric on the cylinder is flat, but the causal structure,
is nevertheless quite bizarre. The lower part of the cylinder is obtained
by identifying causally disconnected pairs of points and, therefore,
causality holds here.  But the upper half originates from the quadrant
$\alpha>0$, $\beta<0$, where we identify causally related points, so
there are closed causal curves in this part of  $M_\text{M}$ (the blue
one, for example, originating from $\gamma$). The boundary between these
regions is formed by the circle obtained from the geodesic $\alpha=0$.
The fate of the other null geodesics of $H$ differ: the vertical ones map
to generators of the cylinder (as, for example, $\lambda$ does). And the
horizontal null geodesics, like $\mu$, say, turn into spirals which
infinitely wind themselves approaching the horizon and never crossing it.
The timelike curves in $H$ which do not cross the ray $\alpha=0$ behave
the same way, see the green curve in Fig.~\ref{fig:misnb2}. Their images
in the Misner space being sandwiched between two null spirals also
infinitely approach the horizon and, correspondingly, have no end points.
So, we have a lot of endless curves with  finite $b$-lengths. Thus the
Misner space is
\emph{singular}. And as the metric is flat the
singularity is quasiregular (even though it is so different from that
considered above).
\end{teor}

\subsubsection{Local extendability}\label{sec:locext}
Let us consider one more singularity.
\kartil{[h]\psfrag{M1}{$M_1$}
\psfrag{M2}{$M_2$}
}{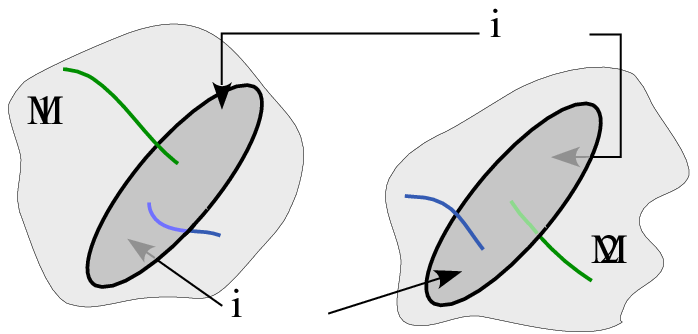}{\label{fig:dpb1}The banks of the slits are identified (which
implies that they lie in isometric domains) so that the green and the
blue lines become continuous. If the isometry can be extended from the
mentioned domains to the entire $M_{1,2}$, the result of the surgery is
the double covering of $M_1-\,$(a sphere of co-dimension 2). }
 Pick two isometric regions (of, perhaps, different spacetimes). Remove
 a closed disk from one of them, the corresponding disk from the other, and
identify the banks of the slits as is shown in Fig.~\ref{fig:dpb1}: the
upper bank of either slit is glued to the lower bank of the other. The
edges of the disks [in the (2+1)-dimensional case these will be circles
$S^1$] cannot be glued back in the space, so the spacetime is singular.
\begin{teor}{Example}\label{ex:DP}
 Two equal spacelike slits separated by time are made in  the Minkowski
 plane. The banks of the slits
 are identified as shown in Fig.~\ref{dpb2}a.
\kartil[\hspace*{4em} (a)\hfill (b)\hspace*{4em}]{[h] \psfrag{u}{$U$}
\psfrag{up}{$U'$}}{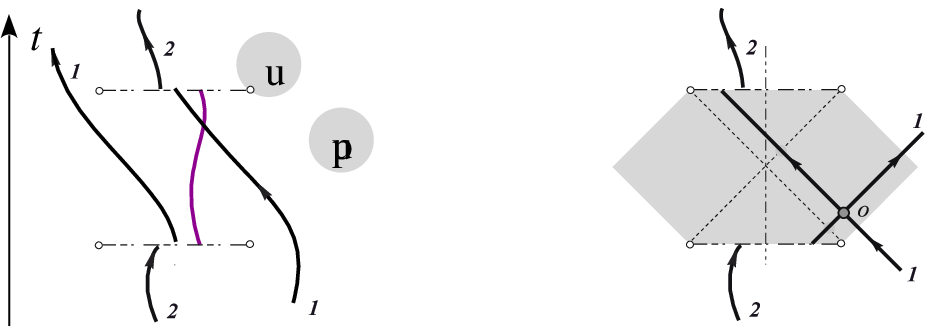}%
{Black curves 1 and 2 are continuous. (a) The purple curve is timelike
and closed. (b) Shadowed is the region where causality does not hold.
\label{dpb2}} The resulting spacetime
\cite{Deu} contains closed causal curves  --- the purple one, for example ---
and is referred to as the Deutsch-Politzer (DP) time machine. Later we
shall use a slightly different spacetime, see Fig.~\ref{fig:dpb1}b,
called the twisted  Deutsch-Politzer (TDP) space. It is obtained the same
way as the DP space, but one of the banks before being glued to the other
is mirror inversed. The TDP space is non-orientable, but this doesn't
matter much, because its 4-dimensional analogue can be made orientable.
What does matter is that a typical null geodesic entering the domain of
causality violation has a self-intersection.
\end{teor}

Both spacetimes in Example~\ref{ex:DP} have singularities. However, if we
had glued the banks in the ``normal'' way we would have been able to
remove the singularities arriving simply at the Minkowski space. To
capture this idea Hawking and Ellis introduced \cite{HawEl} the concept
of local extendability:
\begin{teor}{Definition}
A spacetime $M$ is said to be \emph{locally extendible} if it contains an
open set  $U$ such that (i) the closure of  $U$ is non-compact, but (ii)
$U$ is isometric to a subset $U'$ of a spacetime $M'$ in which Cl$\,U'$
is compact.
\end{teor}
Loosely speaking, locally extendible spaces are those where some points
are missing which  ``might have'' existed. The DP space is obviously
locally extendible (for example, the closure of the ball $U$, see
Fig.~\ref{fig:dpb1}a, is evidently non-compact, while the closure of
$U'$, which is isometric to $U$, is compact) and so is $\EuScript M$.

\subsubsection{Holes}

The concept of local extendability operates with a spacetime as a whole,
with no reference to anything like evolution. There is an alternative
concept, however, formulated in causal rather than in topological terms.

First, for a given set $\EuScript S$ we define the domain of dependence
$D^+(\EuScript S)$ to be the collection of all points $p$ such that every
past endless curve through $p$ meets $\EuScript S$.
Let, for example, $\EuScript S$ be a line of constant time in the
Minkowski plane. Its domain of dependence is the whole upper half-plane.
At the same time in the cone   $\EuScript M$   the domain of dependence
of the ``same'' line does not include the pale purple region, see
Fig.~\ref{fig:primerb}a. Indeed, through any point of this region there
is a past endless curve (like that drawn in brown) which does not meet
$\EuScript S$. Such curves appear, of course, due to the ``missing''
points and hence the definition\footnote{On the (minor) deviations of
this definition from the original one \cite{GerHole}, see
\cite{hole}.}.

\begin{teor}{Definition}
A space-time $(M, g)$ is called
\emph{hole-free}  if it has the following property: given
any achronal\footnote{A set is \emph{achronal} if no its points can be
connected with a timelike curve.} hypersurface $\EuScript S$ in $M$ and
any metric preserving embedding $\pi$ of an open neighbourhood of
$D^+(\EuScript S)$ into some other spacetime $(M',g')$, then
$\pi(D^+(\EuScript S))= D^+(\pi(\EuScript S))$.
\end{teor}
In other words, a hole-free space is that where the domain of dependence
of any surface is ``as big as possible''. Clearly, the spacetime
$\EuScript M$ is not hole-free.

Summing up,  the  singularity in $\EuScript M$ is a quasiregular
singularity. At the same time $\EuScript M$ is a locally extendible
spacetime. And, finally, $\EuScript M$ is is not hole-free. Which  of
these properties should be forbidden?

\subsection{Impasses}
\subsubsection{The ``unphysical nature'' of quasiregular singularities}
In their pioneering paper on quasiregular singularities
\cite{quasireg} Ellis and Schmidt speaking through Salviati
say: ``We know lots of examples of quasiregular singularities, all
constructed by cutting and gluing together decent space-times; and
because of this construction, we know that these examples are not
physically relevant.'' But this argument --- in spite of its popularity
--- is emphatically untenable: the cuttings and gluings that we used are
not
\emph{a property} of the relevant spacetimes, they only are a means of
description.
\emph{Any} spacetime can be constructed by cutting and gluing together
some other decent spacetimes and \emph{any} of them can be constructed
otherwise. The spacetimes with the singularities in discussion are
absolutely no different in this respect from the others.

Yet another idea was to take into consideration the  Einstein equations
and to show that  quasi\-regular singularities  appear only in spacetimes
with the stress-energy tensor of a very special type. Then matter of that
type could have been declared unphysical and the whole problem would have
been solved. This program, however, does not work. At least, not in the
general case. Indeed, pick an arbitrary spacetime $M_1$ and define $M$ to
be the double covering of  $M_1-\,$(a sphere of co-dimension 2). $M$ can
be visualized as the result of the surgery described in the beginning of
section~\ref{sec:locext}.
The missing sphere gives rise to a quasiregular singularity in $M$, but
since the spacetime $M_1$ was chosen arbitrarily,  $M$ can be built
obeying the Einstein equations with an arbitrarily nice right-hand side.
Moreover, the same would be true for
\emph{any} local condition which one could impose on spacetimes. So, we
conclude that generally
\begin{quotation}
No local condition can exclude quasiregular singularities.
\end{quotation}

\subsubsection{The local extendability postulate}
Maybe then we should forbid a realistic spacetime to be locally
extendible\footnote{In spite of its name, this property is not local.}
\cite{HawEl}? Indeed, the singularities considered in Section~\ref{sec:locext}
appeared, as Hawking and Ellis put it, only because we were perverse
enough to extend the top and bottom sides of the slits ``wrong way''. So,
if we consider a four-dimensional spacetime as a result of some
``evolution'' (the evolution, say, of a three-dimensional space with
time), such a postulate seems self-suggesting. What it says is, loosely
speaking, the following. In its evolution a spacetime at every moment of
time has to choose between   developing a quasiregular singularity and
avoiding it, and the spacetime always choose the latter. However, as Beem
and Ehrlich showed
\cite{BeE}, this approach does not work either.

Consider
\begin{figure}[h,b,t]
\begin{center}
\psfrag{t}{$t$}
\psfrag{x}{$x$}
\psfrag{tp}{$t'$}
\psfrag{xp}{$x'$}
\psfrag{m}{$\pi$}
\psfrag{me}{$\rmd s^2=\rmd x^2 -\rmd t^2$}
\psfrag{mec}{$\rmd s^2=\rmd x'^2 -\rmd t'^2$}
\psfrag{U}{$U$}
\psfrag{Up}{$\pi(U)$}
\includegraphics[width=\textwidth]{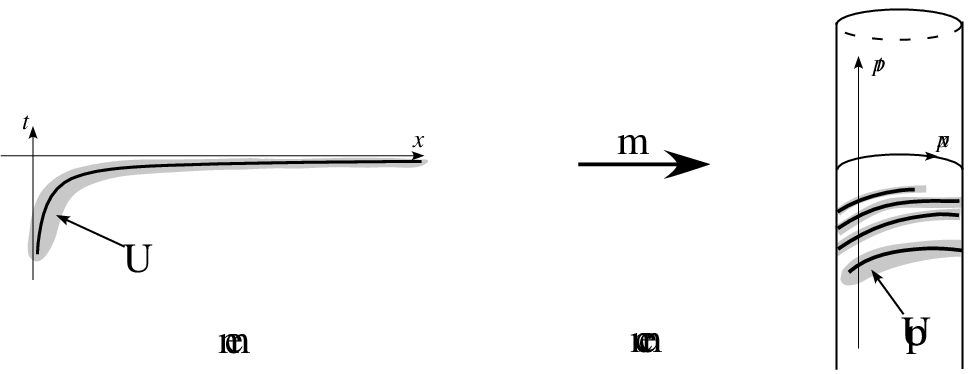}%
\end{center}
\caption{\label{beem_ehr}The Minkowski plane is locally extendible.}
\end{figure}
a flat cylinder with the period $l$ and a map $\pi$ sending each point of
the Minkowski plane to a point of the cylinder according to the rule
$$t'=t,\qquad x'=x \mod l$$ (we wrap the plane around the cylinder, see
Fig.~\ref{beem_ehr}). The part of the hyperbolae shown in
Fig.~\ref{beem_ehr} is mapped to a spiral: a curve which infinitely
approaches the circle $t'=0$ and has no self-intersections. A
neighbourhood $U$ of the hyperbolae, if chosen appropriately (i.~e., to
be narrow enough), will map to its image one-to-one and, obviously,
isometrically. So, the restriction of $\pi$ to $U$ is a metric preserving
embedding. And, nevertheless, the closure of $\pi(U)$ is compact, while
the closure of $U$ is not. Thus   we conclude that
\begin{quotation}
the Minkowski space is locally extendible.
\end{quotation}
But a postulate forbidding even the Minkowski space is definitely
\emph{too} strong.

\subsubsection{Hole-freeness}
The last possibility\footnote{Of those considered in this lecture.} is to
use ``hole-freeness'' as a criterion and, following Geroch's proposal, to
``modify general relativity as follows: the new theory is to be general
relativity, but with the additional condition that only hole-free
spacetimes are permitted'' \cite{GerHole}. The proposal seems to be
physically well motivated. Indeed, what happens in the domain of
dependence of $\EuScript S$ is causally determined by the data on
$\EuScript S$, so one can easily imagine that the existence itself of a
point of $D^+(\EuScript S)$ is also determined by them and not by the
remainder of the spacetime.

However, this approach fails by exactly the same reason as the previous
one: even the Minkowski plane is not hole-free.

\begin{figure}[htb]\begin{center}
\psfrag{S}{$\EuScript S$}
\psfrag{D}{$D^+(\EuScript S)$}
\psfrag{d}{$\shir(t_2)$}
\psfrag{T}{$\tau$}
\psfrag{c}{$\chi$}
\psfrag{p}{$\pi$}
\psfrag{R}{$R'$}
\psfrag{g}{$\gamma$}
\psfrag{g1}{$\gamma^\infty$}
\psfrag{U}{$U$}
\psfrag{Mp}{$M'$}
\psfrag{t}{$t$}
\psfrag{t1}{$t_1$}
\psfrag{t2}{$t_2$}
\psfrag{t3}{$t_3$}
\psfrag{x}{$x$}
\psfrag{Sp}{$\pi(\EuScript S)$}
\includegraphics[width= 0.8\textwidth]{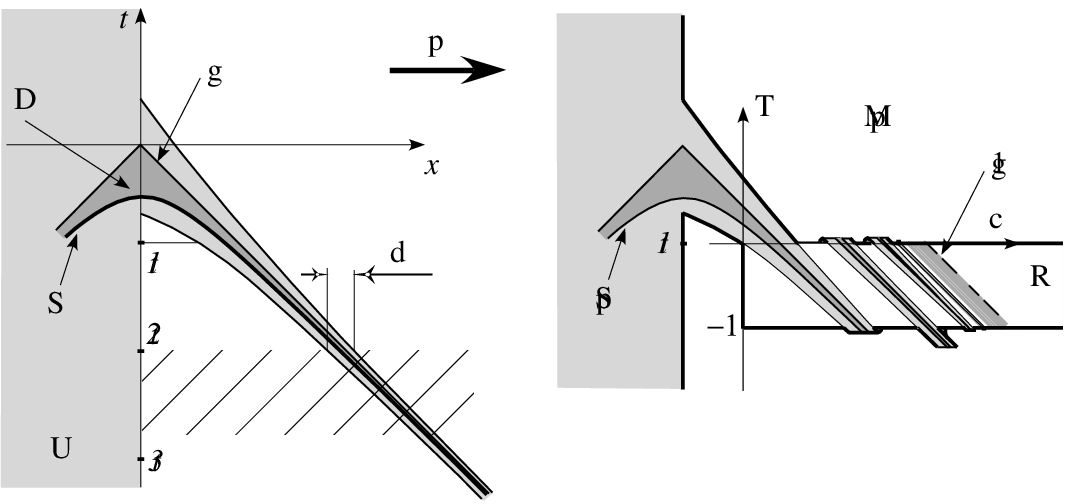}
\end{center}
\caption{\label{fig: ahb} The dark-gray regions are $D^+(\EuScript S)$
and its image. By light-gray $U$ and $\pi(U)$ are shown.}
\end{figure}

\paragraph{\it Proof.} Let $\EuScript S$ be a  hyperbolae in the Minkowski
plane. Its domain of dependence $D^+(\EuScript S)$ is the closed set
shown in Fig.~\ref{fig: ahb}; it is bounded (in particular) by a past
directed null geodesic $\gamma$.  The neighbourhood $U$ of $D^+(\EuScript
S)$ is defined to be the left half-plane plus a ``beak'', see
Fig.~\ref{fig: ahb}. The beak is characterized by width $\shir$ and is
chosen so that
\begin{equation*}
\shir \text{ is monotone and } \shir (t) \to 0  \text{ at }  t\to-\infty.
\end{equation*}
The idea is to find for $U$ an alternative  extension (not the Minkowski
plane, but some $M'$ instead) and to check that in that extension the
domain of dependence of the image of $\EuScript S$ is larger than simply
the image of $D^+(\EuScript S)$:
\begin{equation}\tag{$*$}\label{eq:defhole}
D^+(\pi(\EuScript S))\varsupsetneq
\pi(D^+(\EuScript S)).
\end{equation}

The extension $M'$ is built by gluing some portions  $O_k$, $k=1,\dots$
of the beak to a rectangular strip $R$:
\[
\rmd s^2= -\rmd \tau^2 +\rmd\chi^2, \qquad
\tau \in (-1,0),\quad\chi>0.
\]
Specifically,  pick a sequence $\{t_k\}$ of negative numbers such that
$t_{k+1} < t_{k} -1$ and define  $O_k$ as follows:
\[
O_k\equiv\{p\in U\colon\quad x(p)>0,\quad t_{k}>t(p)>t_{k}-1\}
\]
(for example, the hatched strip in Fig.~\ref{fig: ahb} cuts $O_2$ out of
the beak). Further, let $\Psi$ be the isometry which sends, for every
$k$, each point $p\in O_k$ to the point $q\in R$ according to the rule
$$\tau(q)= t(p) - t_k,\qquad
\chi(q)= x(p)-\chi_k
$$
Now $M'$ is defined as the quotient:
\[
M'\equiv U\cup_\Psi R
\]
(i.~e., as the result of gluing together $U$ and $R$ by $O$) and $\pi$ as
the natural projection of $U$ to $M'$.

I have not specified $\chi_k$ (see \cite{hole} for details). but, in
fact, all required of them is that $\pi(O_k)$ would not not overlap and
at the same time the series $\sum
\shir_k$ would converge, where $\shir_k$ is the maximal width of $O_k$ (for
$\{t_k\}$ falling fast enough such a choice of $\chi_k$ is obviously
possible). The above-mentioned convergence implies that there is a null
geodesic segment $\gamma^\infty$
--- the dashed line in Fig.~\ref{fig: ahb} --- to which the null geodesics
$\pi(\gamma\cap O_k)$ converge.

It is the existence of  $\gamma^\infty$ that proves \eqref{eq:defhole}
and hence the proposition. Indeed, $\gamma^\infty$ is disjoint with
$\pi(D^+(\EuScript S))$. But, on the other hand, a timelike curve through
any its point meets inevitably $\pi(\EuScript S)$. So,
$\gamma^\infty\subset D^+(\pi(\EuScript S))$. \\
\hspace*{1cm}\hfill $\Box$

\section{Formation of quasiregular singularities}

As we have just seen there is no obvious way to expel quasiregular
singularities from relativity. Now let us formulate the question the
other way round: Are there any indications that quasiregular
singularities do exist, or may appear under favorable conditions? At
first glance the property which makes such singularities interesting ---
their ``suddenness'' --- makes
 at the same time their appearance unlikely. Indeed, if everything
prior to the singularity looks \emph{as if} the spacetime is
singularity-free, why not assume that it \emph{is} --- and will remain
--- singularity-free. Actually, however, the situation is somehow less
trivial. I think, two possibilities are of interest here. Let us start
with the more subtle and academic one.
\subsection{How to artificially create a
quasiregular singularity}

\subsubsection{Time machines}
In our consideration we have already met two spacetimes --- the DP space
and the Misner space ---  which evolve nicely up to some moment, but then
lose causality; in other words closed causal curves appear to the future
of some globally hyperbolic domain. The regions where causality breaks
down and the entire spacetimes containing them are called \emph{time
machines}.
\begin{figure}[h,b,t]
\begin{center}
\psfrag{D}{DP spacetime}
\psfrag{TD}{Twisted DP spacetime}
\psfrag{M}{Misner space}
\psfrag{p}{$p$}
\psfrag{ct}{$t$}
\psfrag{t}{$\tau(p)$}\psfrag{r}{$\tau(p')=\tau(p)$}
\psfrag{pp}{$p'$}
\includegraphics[width=0.28\textwidth]{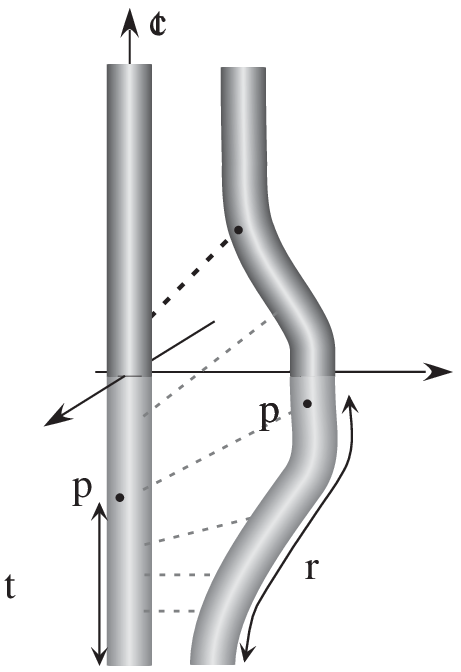}%
\end{center}
\caption{Wormhole-based time machine\label{fig:fabric}. The lengths are
measured from a surface $t=const$. The thick dashed line is the earliest
closed causal (null) curve.}
\end{figure}
 The most known time machine  is probably that based on a
wormhole
\cite{MTY}. The relevant spacetime, see Fig.~\ref{fig:fabric}, is obtained, for
example, by removing\footnote{Again, this surgery is nothing more than a
description. No real cutting of a spacetime is meant.} two timelike tubes
from the Minkowski space and identifying the boundaries of the holes (the
vicinity of the junction is understood to be smoothed out appropriately,
so that the junction is seamless).  The identification is done as
follows: every generator of the left cylinder is glued to the
corresponding generator of the right cylinder so that the length of the
corresponding segments are   equal. In other words, the clocks traveling
with one mouth of the wormhole are synchronized (when seen through the
throat) with those left at rest near the other mouth. Clearly, if the
right tube is tortuous enough, then the points which are to be identified
become causally separated beyond some surface. So, closed causal curves
appear to the future of a globally hyperbolic region and the spacetime
becomes a time machine.

At first glance the mentioned time machines form two --- fundamentally
different --- classes. The DP time machine, as we discussed, may appear
or may not. One can neither predict its appearance, nor cause it. And it
seems to be completely different with the wormhole-based time machine or
the Misner space, where one prepares suitable initial conditions and
waits until the spacetime governed by the Einstein equations gives birth
to a closed causal curve. So, one can even think that time machines are
divided into spontaneously appearing and artificially manufactured. In
fact, however, this difference is a sheer illusion. \emph{Any} spacetime
in its evolution can avoid transformation into a time machine. The
following theorem is proved

\paragraph{Theorem \cite{teorema}.} Any spacetime $U$ has a maximal
extension $M^{\rm max}$ such that all closed causal curves in $  M^{\rm
max}$ (if they exist there) are confined to the chronological past of
$U$.

\begin{figure}[h,t,b]
\begin{center}
\psfrag{t}{$t$}
\psfrag{U}{$U$}\psfrag{cuts}{cuts}
\psfrag{a}{$\alpha$}\psfrag{b}{$\beta$}
\psfrag{cuts}{cuts}
\includegraphics[width=0.6\textwidth]{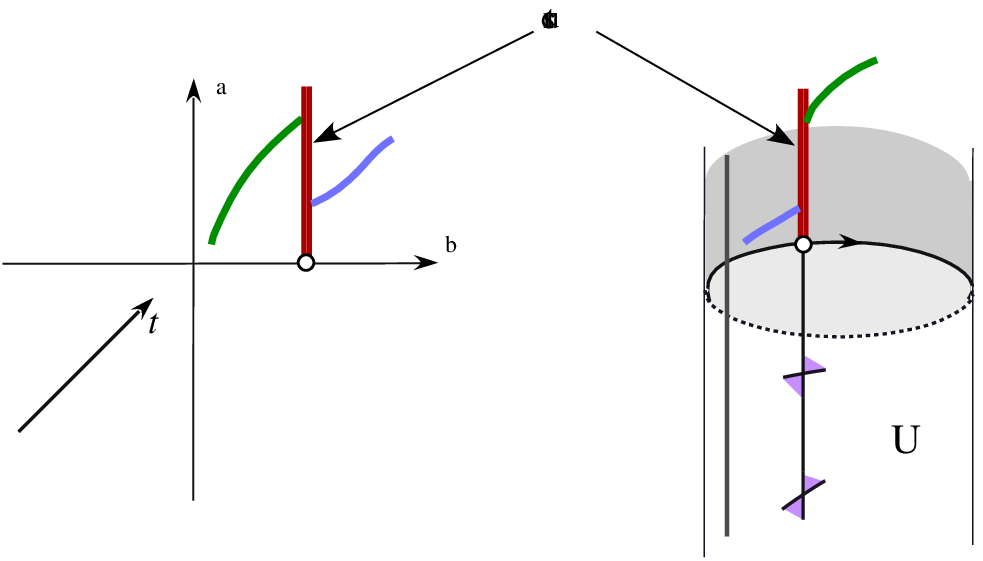}%
\end{center}
\caption{\label{fig:disarm} $U$ is the lower (causality-respecting) half
of the Misner space. The circles designate a ``new'' quasiregular
singularity.}
\end{figure}

The proof is quite lengthy, so I shall only illustrate the theorem by the
example of the Misner space. Denote by $U$ its initial, globally
hyperbolic part. An inhabitant of $U$ sees that in the course of time the
null cones open more and more, see Fig.~\ref{fig:disarm}.
So, he anticipates the unavoidable (as he might think, knowing from the
Einstein equations that the spacetime will have to remain flat)
appearance of a time machine (when the inclined generator becomes
horizontal). What the theorem says is that his hopes may be vain: besides
the Misner space, $U$ also has another,
\emph{causality-respecting} (and also flat) maximal extension. To
verify this assertion recall that in the Misner space  the vertical lines
generating the cylinder are null geodesics. So, there is a neighbourhood
of the brown, say, ray isometric to a neighbourhood of a null ray of the
Minkowski plane. The desired extension now can be described as the result
of removing these two rays and gluing the right bank of either cut to the
left bank of the other (so that  the green and blue curves in
Fig.~\ref{fig:disarm} are continuous). That causality holds in the thus
constructed extension is obvious. Note, however, that this is achieved at
the cost of allowing a new quasiregular singularity to appear.

It's important that by ``spacetime'' in the theorem not just a smooth
connected Hausdorff pseudo-Euclidean manifold is understood. One can
impose \emph{any} additional condition and, as long as it is local (like
the Einstein equations), the theorem remains true \cite{teorema}.

Summing up, whatever we do within general relativity we cannot force the
universe to give birth to a time machine. What we \emph{can} do, however,
is force the universe to \emph{choose} between creating a time machine,
or a quasiregular singularity. And there are indications as I shall show
in a moment that the universe might prefer the latter.
\subsubsection{Time travel paradox}
Consider the following situation. An experimenter learns how to build  a
time machine (out of a wormhole, say). He makes all necessary
preparations to ensure that it will appear in 5 minutes.
\begin{figure}[h,b,t]
\begin{center}
\includegraphics[width=0.8\textwidth]{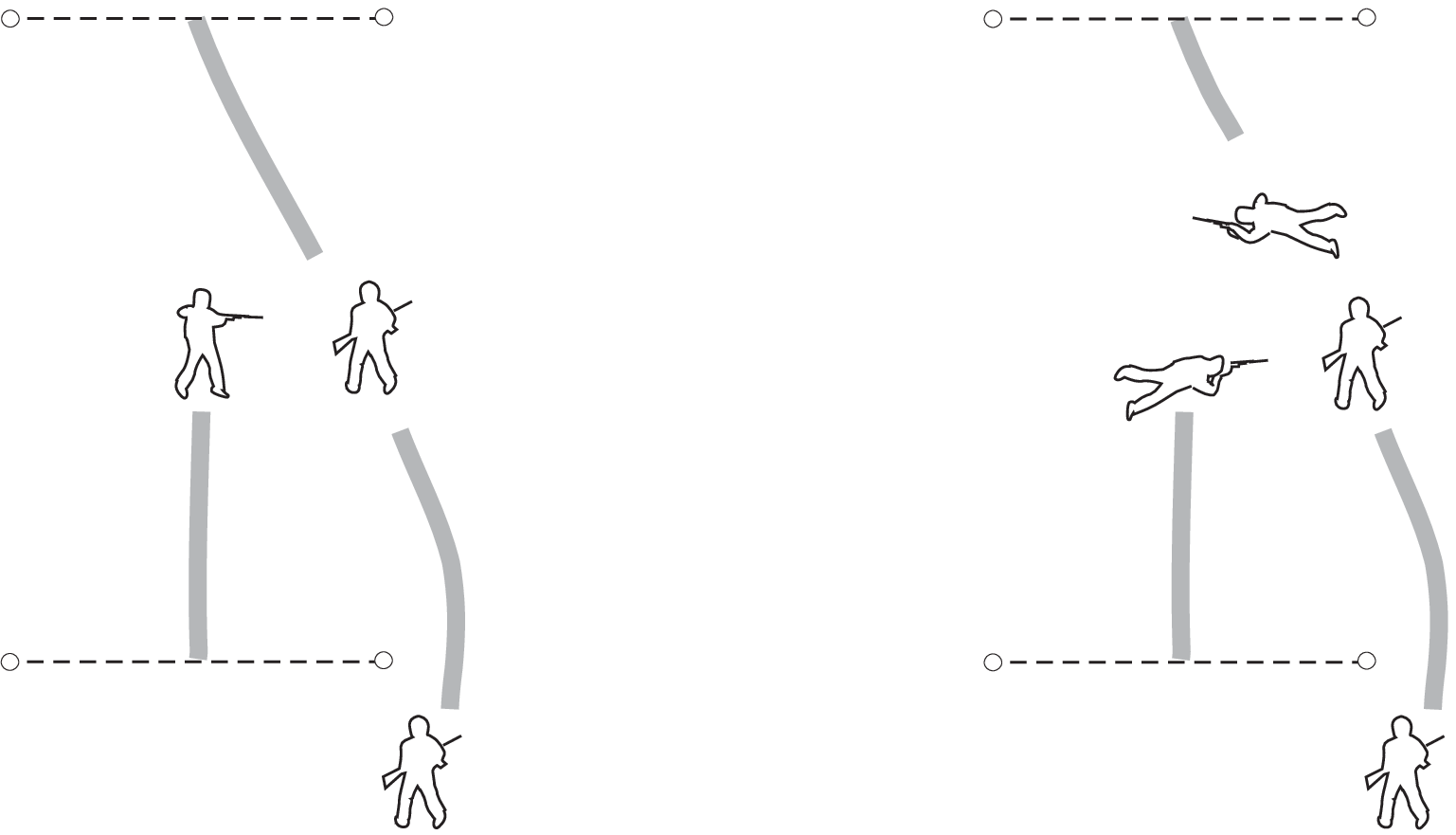}
\end{center}
\hspace*{8em} (a)\hfill (b)\hspace*{8em}
\caption{\label{fig:suic} (a) An apparent time travel paradox. (b) One of
its possible resolutions.}
\end{figure}
Than he loads his rifle and makes a firm decision
 to enter the time machine as soon as it appears, and to return with its
 help ``back in time'', where to
waylay his younger self, see Fig.~\ref{fig:suic}a, and  shoot the latter
dead. It is clear that his plan cannot be realized. What is
\emph{not} obvious is what actually will happen.

Let us reformulate the situation in more general terms. We have a system
in some initial state (an armed person in a room where a time machine is
to appear). The system is governed by some quite plausible laws of motion
(the person must wait for 5 minutes then make a few steps, raise the
rifle, and pull the trigger). There is nothing pathological either in the
initial state, or in the laws of motion. In a globally hyperbolic
spacetime they would uniquely determine the evolution of the system. But,
because the spacetime contains a time machine, we are facing a
paradox\footnote{Not to be confused with the ``grandfather paradox'',
which is, strictly speaking, not a paradox at all \cite{parad}.}: no
evolution corresponds to the specified combination of the initial state
and laws of motion. How should this be interpreted?

A possible solution is this. The system in consideration is highly
complex and actually we did not specify its laws of motions accurately
enough. So, it might happen that we are just overlooking the solution.
For example, one can advocate the evolution depicted in
Fig.~\ref{fig:suic}b: the experimenter is wounded (not killed!) and it is
the wound that in due time prevents him from shooting accurately and thus
leads to wounding the target instead of killing it.

\begin{figure}[h,b,t]
\begin{center}
\psfrag{O}{$O$}
\psfrag{t}{$t$}
\psfrag{t0}{$t_0$}
\psfrag{t1}{$t_1$}
\includegraphics[width=0.8\textwidth]{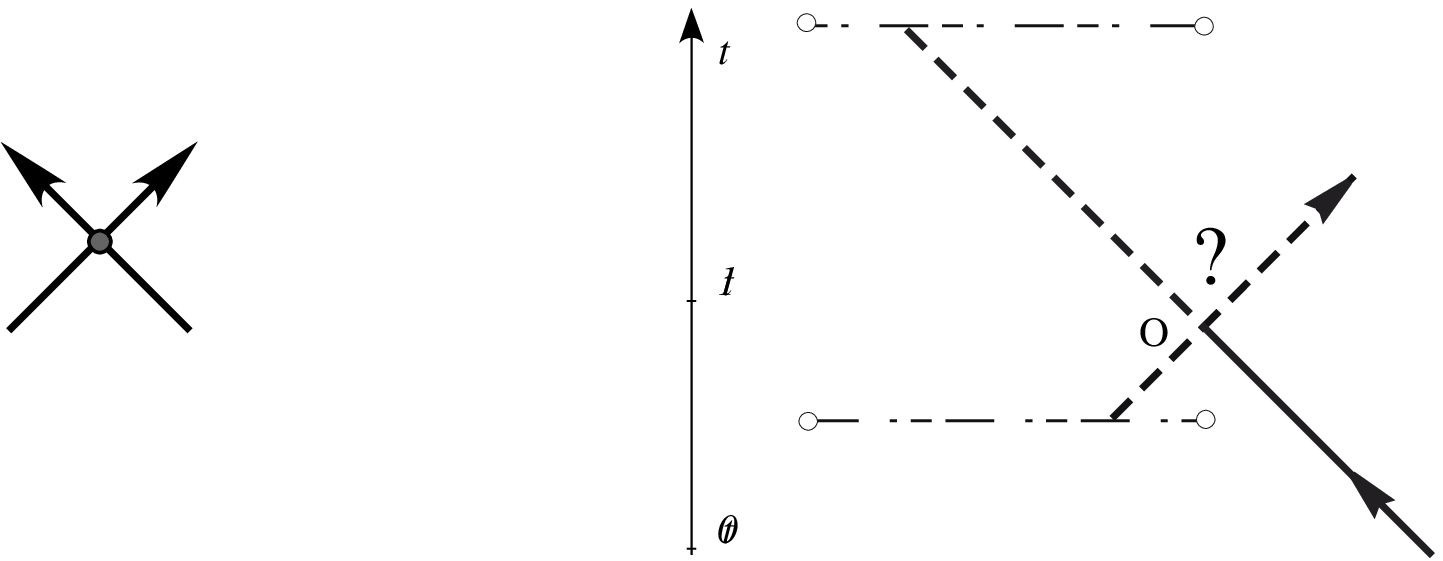}
\end{center}
\hspace*{6em} (a)\hfill (b)\hspace*{6em}
\caption{(a) The vertex defining  dynamics in our toy model.
(b) The ``paradox'' appearing if the number of particles assumed to be
conserved.
\label{fig:tpseudopar}}
\end{figure}

Such reasonings may lead one to the following
\begin{teor}{Conjecture}\label{conj}
Any reasonable laws of motion being combined with any reasonable initial
state must correspond to some evolution whether or not a time machine
appears.
\end{teor}
To cast doubt on this conjecture and to construct a time travel paradox
\cite{parad} it is instructive to analyze a toy model of this situation.
Consider the twisted DP universe, see example~\ref{ex:DP},  populated
only by pointlike particles, whose evolution is determined by the
following simple laws.
\begin{enumerate}
  \item[($a$) ] Particles move along null geodesics;
  \item[($b$) ] They cannot appear from nothing or disappear (local
  conservation);
  \item[($c$) ] Their interaction reduces to the vertex shown in
  Fig.~\ref{fig:tpseudopar}a.
\end{enumerate}
At first glance it might seem that already the initial data shown in
Fig.~\ref{fig:tpseudopar}b --- a single particle ready to fly into the
time machine --- constitute the desired paradox.
Indeed, in the TDP space  all null geodesics entering the time machine
have self-intersections. So, if the particle enters the time machine it
must hit its younger self in the point $O$ thus preventing the latter
from getting into the time machine. But if this younger particle does
change its trajectory and fly away, then where  the older particle (the
second participant of the collision) came from? A paradox, apparently.

The entire reasoning, however, is based on the false implicit assumption
that the number of particles is a conserved quantity. So, if there was a
single particle on the surface $t=t_0$, then there must be only one
particle on the surface  $t=t_1$ too. But this global conservation
doesn't follow from our laws ($a$--$c$) and no reasons at all are seen to
impose it as a separate additional law. And as soon as we abandon it, a
nice solution appears satisfying all those laws: it contains
\emph{two} different particles. One of them has a closed world line. The
other collides with that former one in $O$ and bounces away from the time
machine.

To exclude such a solution we can sophisticate the model. Let now the
particles be of two different kinds: dark and light and the interaction
is defined by this table
\begin{center}    \includegraphics[width=0.4\textwidth]{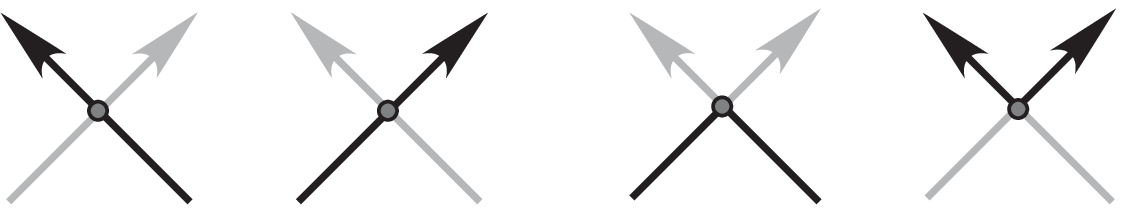}
 \end{center}
The laws look quite realistic in the sense that they are local and
respect all the symmetries of the problem. And again at first glance it
seems that there is no evolution from the  initial data of
Fig.~\ref{fig:tpseudopar}b: one cannot assign in a consistent way a tint
to the closed world line in the figure. Indeed, it is a single particle,
so it must have the same tint all along, but according to the laws that
we have adopted the incoming left and the outgoing left particles always
have different tints.

\begin{figure}[h,t,b]
\begin{center}
\psfrag{O}{$O$}
\psfrag{OP}{$O'$}
\includegraphics[width=0.4\textwidth]{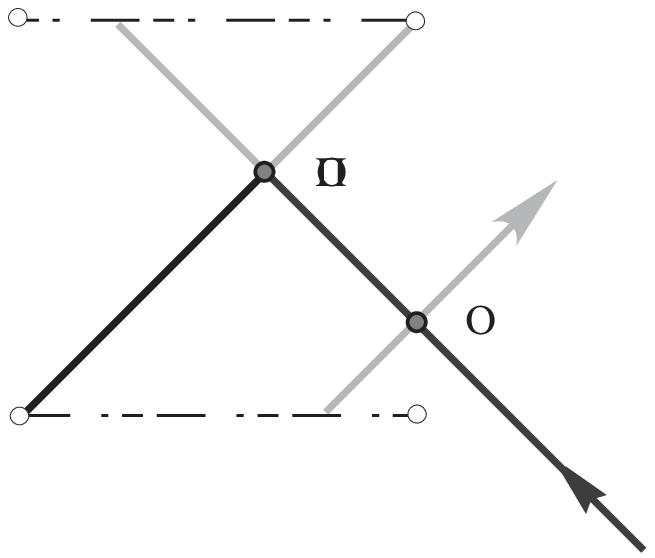}
\end{center}
\caption{\label{fig:uslozhcvx} There are three light particles in this world:
one is emitted in  $O$ and escapes to infinity and two others are born in
the collision at $O'$. Of them one is absorbed in   $O$ and the other
disappears in the singularity.}
\end{figure}

And still this is not a paradox yet. In Fig.~\ref{fig:uslozhcvx} we see
one of the admissible solutions.
It  contains three dark and three light particles. The former include the
particle coming from infinity to the point $O$, another one, emitted in
$O$ and absorbed in $O'$, and, finally, the particle emerging from the
singularity to be absorbed in $O'$. This is a legitimate evolution
satisfying all the laws formulated above.

\begin{figure}[h,t,b]
\begin{center}
\psfrag{h}{$h$}
\psfrag{t}{$t=t_0$}
\includegraphics[width=0.4\textwidth]{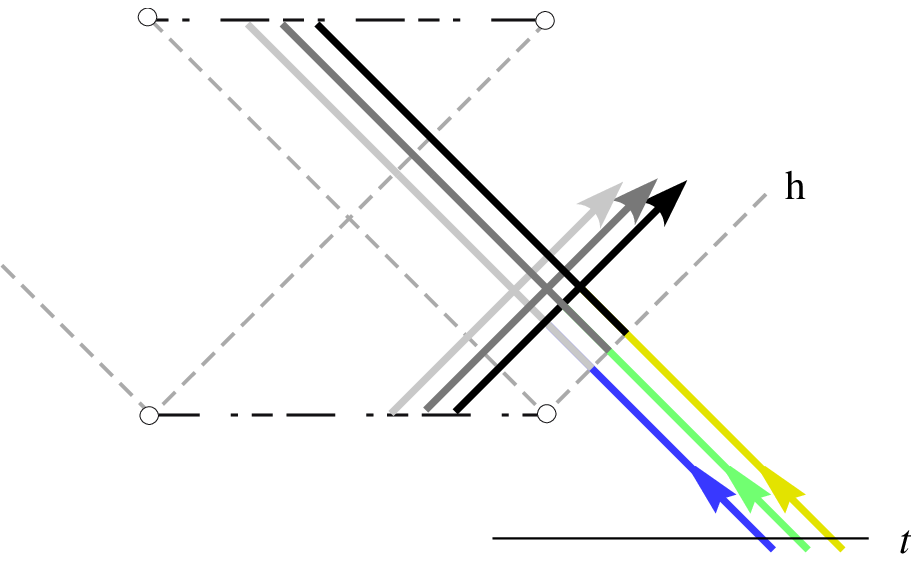}
\end{center}
\caption{\label{fig:nparadob} Any evolution must include the solid
lines as world lines of particles. It also may include some of the dashed
lines, but nothing else.}
\end{figure}

Now we have discovered all pitfalls and the next sophistication brings us
the desired paradox \cite{parad}. Namely, suppose that every particle in
the world under discussion has one more characteristic, let us call it
color. There are three different colors and the particles of different
colors do not interact. Then the initial state shown in
Fig.~\ref{fig:nparadob} does not
correspond to any evolution. Indeed, any solution must have three
self-intersecting world lines to the future of $h$. Only two of them may
be affected by collisions with particles of the same color (the possible
world lines of such particles are shown by dashed lines). So, the tint of
at least one particle with the self-intersecting world line remains
constant, which, as discussed above, contradicts the laws of interaction.

Of course, this model is \emph{very} simple. Nevertheless, it strongly
suggests that Conjecture~\ref{conj} is false and in some cases the
configuration of the matter fields may exclude the appearance of a time
machine. If so, we come to a rather unexpected conclusion:
\begin{quotation}
 By arranging matter particles one can make some of geometrically
 admissible extensions of a given spacetime impossible.
\end{quotation}
In other words, the matter content of our world determines its geometry
\emph{not only} via Einstein's equation \cite{parad}.

A pertinent consequence is that all one needs to create a quasiregular
singularity is a wormhole and a rifle. Then forcing the universe  to
choose between the singularity and the time machine one can make at the
same time the appearance of the latter impossible.

\subsection{Primordial quasiregular singularities}
Now let us turn to another relevant environment, which is the early
Universe.
 In this section I briefly remind the reader the arguments suggesting
 that there are cosmic
strings in nature (all details can be found in \cite{SheVi}) and then
reproduce them in a purely gravitational case to show that the existence
of string-like singularities (a variety of quasiregular singularities) is
equally realistic.

\subsubsection{Cosmic strings}
Suppose, after the universe had cooled below some critical temperature, a
complex scalar field $\chi$ appeared
\begin{figure}[h,t,b]
\psfrag{C}{$C$}
\psfrag{V}{$V$}
\psfrag{I}{Im$\, \chi$}
\psfrag{R}{Re$\, \chi$}
\psfrag{ch}{$\chi_0$}
\begin{center}
\includegraphics[width=0.7\textwidth]{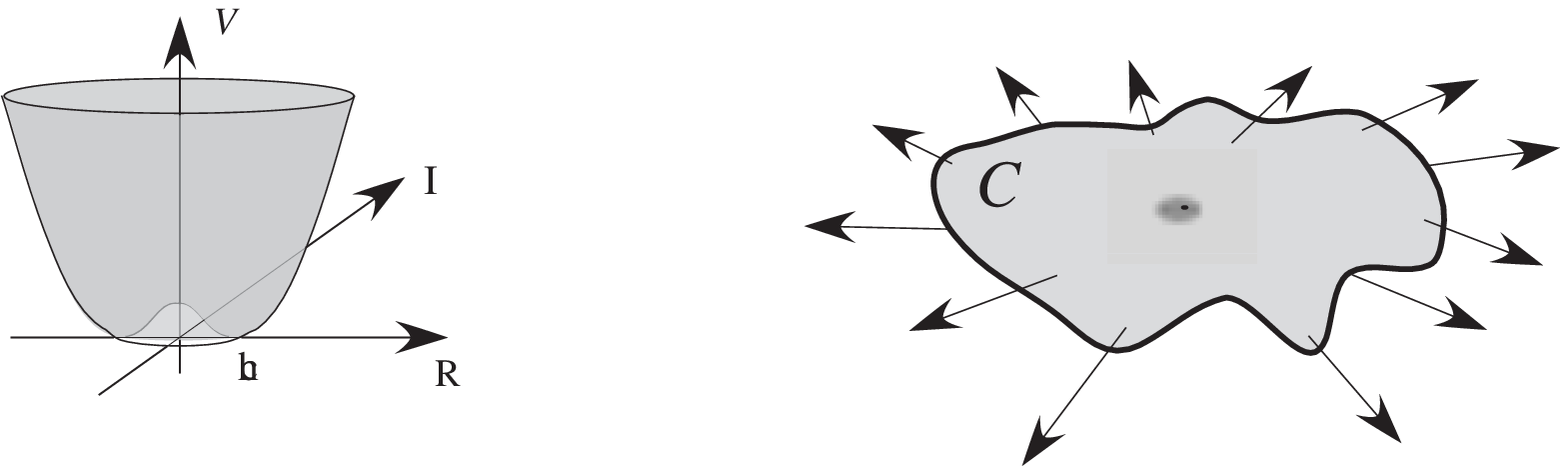}
\end{center}
\hspace*{8em} (a)\hfill (b)\hspace*{8em}
\caption{\label{fig:risb1} (a) The potential has the minimum at
$\chi_0=|\chi_0|e^{i\sigma}$, where $\sigma$ is real, but otherwise
arbitrary. (b) The arrows are complex numbers of the same modulus
$\chi_0$ (the values of $\chi$) and not three-vectors.}
\end{figure}
 with the potential shown in
Fig.~\ref{fig:risb1}a. One expects the field to take the value $\chi_0$
corresponding to the minimal energy. But the evolutions of the field in
different regions are
\emph{uncorrelated} and this can make the process of taking this value by
the field energetically prohibited \emph{globally} even though
\emph{locally} it is energetically favorable. Indeed, suppose on some
loop $C$ the field happened to take  values like those depicted in
Fig.~\ref{fig:risb1}a. Then, evidently, on a  surface enclosed by $C$
there will be a point at which the field vanishes. The energy density in
that point will be, on the contrary, non-zero (because $V(0)\neq 0$). Now
we can deform the disk. The same reasoning will show that there is a
non-vacuum spot on the new surface, too.  So, what we have is actually
not just a single spot of non-zero energy density but an endless curve
(either infinite, or closed), or rather an endless thin tube, because
they must be of finite
--- though small --- thickness. The tubes are stable: even though they
are surrounded by vacuum, they cannot dissolve for the topological
reasons just discussed. It is such tubes that are called  cosmic strings.

An important  thing about cosmic strings is their gravitational fields.
In particular, the universe with a straight cosmic string is believed to
be described ---
\emph{at large $\rho$} --- by the spacetime depicted in
Fig.~\ref{fig:primerb}b, or rather, by its four-dimensional analog
\begin{equation}
\begin{split}
\label{1}
\rmd s^2= -\rmd t^2 + \rmd z^2 + \rmd \rho^2 + \rho^2\rmd \varphi^2,
\\
  t,z\in \mathbb R,\quad \rho>0,\quad \varphi=\varphi+2\pi-d,\quad d\in
  (0,2\pi).
\end{split}
\end{equation}
In other words, at large $\rho$  the straight cosmic string produces the
same gravitational field as the singularity   considered in the
Introduction.

\begin{figure}[h,b,t]
\begin{center}
\includegraphics[width=0.6\textwidth]{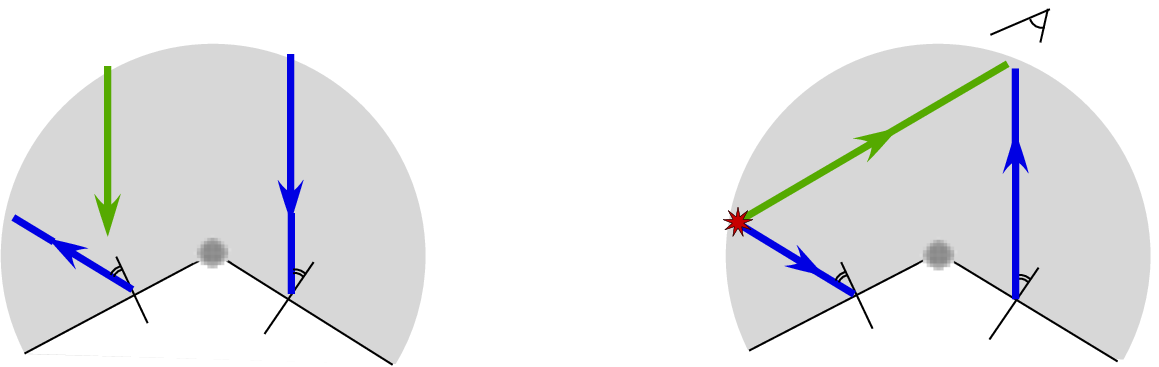}
\end{center}
\hspace*{11em} (a)\hfill (b)\hspace*{11em}
\caption{\label{risb3} Identify the thin line to obtain the section
$t=const$, $z=const$ of the spacetime with a  cosmic string along the
$z$-axis. In fact, such a section is just a flat cone with a smooth cap.}
\end{figure}

When a string moves through the cosmological fluid it leaves a wake
behind it: as is seen from  Fig.~\ref{risb3}a
two \emph{parallelly} moving galaxies may nevertheless collide  after a
string has passed between them --- a phenomenon of obvious importance to
cosmology. On the other hand, two light rays emitted from the same source
may, by exactly the same reasons, come to an observer from different
directions, see  Fig.~\ref{risb3}b. Thus, a string acts as a
gravitational lens producing multiple images of a single object. Note
that both rays propagate in flat spacetime, so the images are neither
distorted, nor fuzzy. They may differ, however, because we see the source
from different angles.

\subsubsection{String-like singularities \cite{grStr}}
Now let us apply all the above to the purely gravitational case. In the
end of the Planck era the classical spacetime had emerged and started to
expand obeying the Einstein equations. By the time it could be
confidently called classical it was practically flat (by Planck's
standards, anyway), so we can speak of the emergence (whatever it means)
of a flat spacetime. One can think, however, that remote regions evolved
uncorrelatedly
\begin{figure}[h,b,t]
\begin{center}
\includegraphics[width=0.6\textwidth]{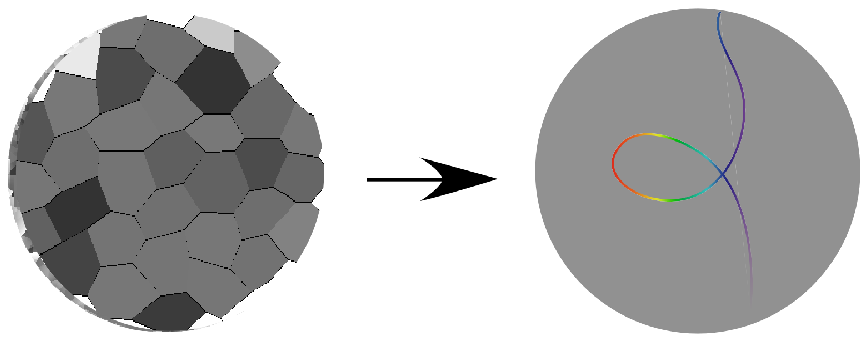}
\end{center}
\caption{\label{fig:risb3}Each patch tends to develop into the Minkowski
space. But as they do so independently, global obstruction may appear
resulting in formation of ``topological defects'' --- string-like
singularities.}
\end{figure}
and again the locally favorable process (of becoming Minkowskian) might
be impeded by some \emph{global} obstructions. Exactly as with matter
fields, such   obstructions must have given rise to   singularities. This
time, however, these would be   true \emph{geometrical} singularities and
exactly of the type we are considering ---  singularities in the
otherwise Minkowski spacetime.

Of course, this scenario contains some handwaving as we know nothing
about how the classical space emerges, but the beauty of it is that we
almost do not need to. Suppose, for example, that a circle lying in a
newborn (non-simply connected) flat region happened to be too short (or
too long) for its radius of curvature. Then we do not, in fact, need to
know anything else to conclude that when the spacetime eventually becomes
entirely classical and flat it will contain a quasiregular singularity
(presumably, it will be similar to the straight-line one considered in
the Introduction and will be represented by an endless --- though not
necessarily straight
--- line). This scenario seems so natural and convincing that I think the
appearance of quasiregular singularities in the Early universe is at
least as realistic as appearance of cosmic strings.
\newpage

\section{Observations}
Now suppose quasiregular singularities  did appear in the Early Universe
and survived to the present day. What do they look like? How do we detect
them?

Unfortunately, there is still neither an exhaustive list of such
singularities, nor even a detailed classification. Almost all we have is
a number of examples, see \cite{strstr} and references therein. One ---
often referred to as ``straight string'' --- is the spacetime
\eqref{1}. A few more \cite{proiastr} are obtained by changing the way in which the two
half-planes mentioned in the Introduction (see, Fig.~\ref{fig:primerb}b)
are identified. If before gluing them together we shift (or boost) one of
them with respect to the other, the properties of the resulting
singularity will change significantly. Suffice it to say that if the
shift is in the $t$-direction the spacetime will contain closed causal
curves. The property shared by all these singularities is that they all
in a sense are straight lines at rest. Recently, however, a number of
singularities in flat spacetime were found \cite{strstr} which are
represented, in the same sense, by curved or moving lines.  All such
singularities including those ``straight'' ones I shall collectively call
\emph{string-like}.

\begin{teor}{Examples}\label{ex}
 A) In the flat space $\rea^3$ consider the surface $H$ given, in the
\psfrag{B1}{$\mathcal B_1$}
\psfrag{B2}{$\mathcal B_2$}
\begin{figure}[hbt]
\psfrag{S}{$\eu S$}
\psfrag{z}{$z$}
\psfrag{r}{$\rho$}
\begin{center}
\includegraphics[width=0.8\textwidth]{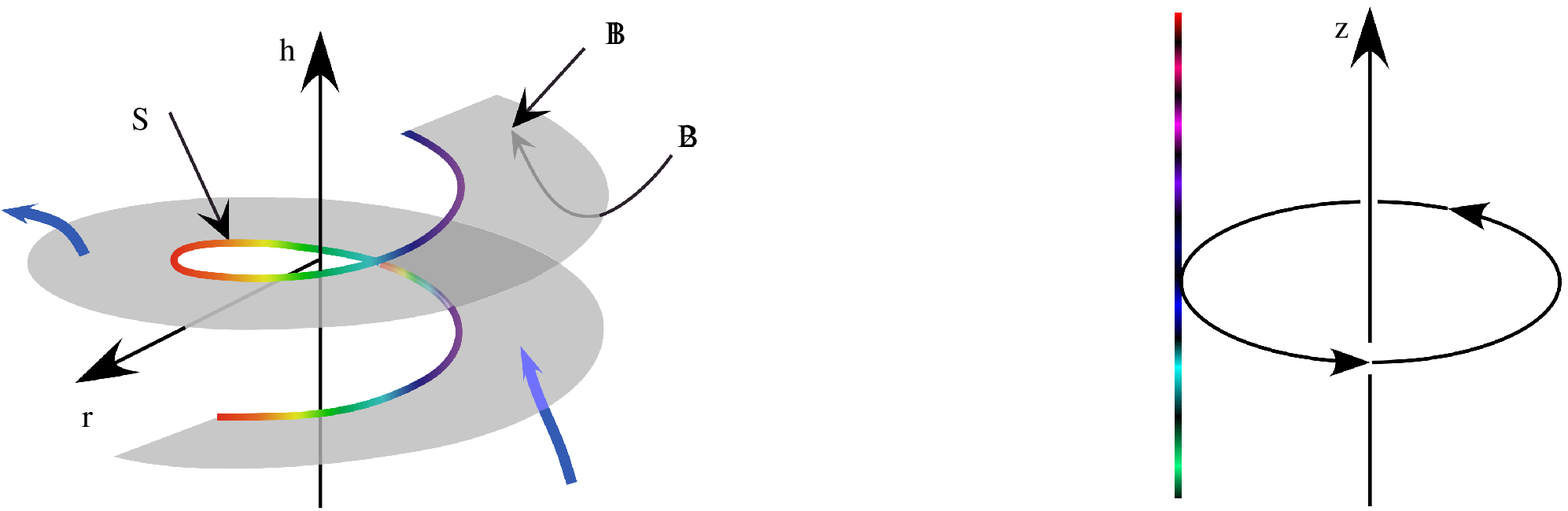}
\end{center}
\hspace*{8em} (a)\hfill (b)\hspace*{8em}
\caption{\label{fig:spiral}(a) The cut is made along the gray surface
(which actually spreads to infinity). (b) The spiral singularity in the
case $\zeta=t$. }
\end{figure}
 cylinder coordinates, by the equation $
 \phi=b\xi\mod2\pi$, where $\rho>\rho_0>0$ and $b\neq 0$.
$H$ is (a half of) a helicoid without the core, see
Fig~\ref{fig:spiral}a,
and is bounded by the spiral
\[
\s\colon\ \qquad \phi=b\xi\mod2\pi,
\qquad \rho= \rho_0.
\]
Make a cut in  $\rea^3$ along Cl$\,H$ (note that $\s$ is removed, too),
rotate the lower bank of the slit
--- it is denoted by $\mathcal B_2$ in the figure
--- anti-clockwise shifting it at the same time upward so that $\mathcal
B_1$ slides over $\mathcal B_2$, and paste  the banks together again into
a single surface. The resulting space $\Mf^3$ is  smooth, flat, etc., but
it lacks the points that formed $\s$ (these points cannot be returned
back insofar as the metric is required to be smooth). Thus, $\Mf^3$ has a
quasiregular singularity and this singularity, when $\xi=z$, has the form
of a spiral (and is called, accordingly,
\emph{spiral}
\cite{strstr}).
 The structure of this singularity becomes more transparent when  $\Mf^3$
 is depicted in the  original
coordinates $z$, $\rho$, $\phi$ as in   Fig.~\ref{fig:spiral}a. These
coordinates are invalid, of course, on $\mathcal B_{1,2}$, that is why a
smooth curve looks discontinuous in the picture. Another curious
singularity is obtained if in the previous procedure one sets $\xi=t$
(and the full spacetime is obtained by multiplying  $\Mf^3$ by the
$z$-axis), see
 Fig.~\ref{fig:spiral}b. It is easy to see that the singularity in this case is
 represented by a
straight line moving  in quite a bizarre manner: it \emph{circles around
nothing}.\\ B) From
\begin{figure}[hbt]
\psfrag{a}{$\alpha$}
\psfrag{E}{$\eucl^3$}
\psfrag{D}{$D$}
\begin{center}
\includegraphics[width=0.7\textwidth]{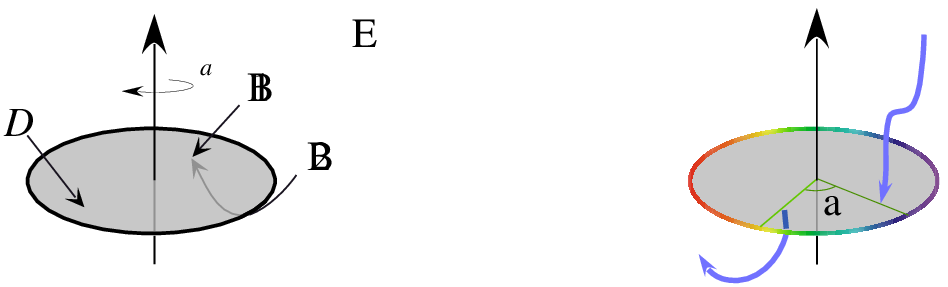}
\end{center}
\hspace*{9em} (a)\hfill (b)\hspace*{9em}
\caption{\label{loop}The \emph{loop} singularity can be view as a special
case of the spiral one.}
\end{figure}
the ordinary 3-dimensional  Euclidean space $\eucl^3$ remove a closed
circle $D$. Then rotate one of the banks ($\mathcal B_1$ in
Fig.~\ref{loop}a) of the thus obtained slit w.~r.~t.\ the other
($\mathcal B_2$, correspondingly) by some $\alpha$, and, finally, glue
the banks together. The resulting space is shown in Fig.~\ref{loop}b,
where as usual we use the old coordinates (so, the blue curve is actually
continuous). The missing circumference $\s$ (as before, it cannot be
glued back into the space) is a closed string-like singularity called
\emph{loop}. It is convenient to think of the spacetime as the Euclidean
space minus the circumference plus the rule that a curve meeting the disk
in some point $p$ is continued from the point obtained from $p$ by
rotation by $\alpha$.
\end{teor}

The relation between the string-like singularities and strings is
summarized in the following  table.
 \begin{table}[hbt]\small
\begin{center}   \begin{tabular}{l|p{0.3\textwidth}||p{0.3\textwidth}}
&\hfil  \bf Cosmic strings   \hfil    \hfil &  \hfil\bf  String-like
singularities   \hfil \\ \hline &&\\
 Source &\hfil  Matter  \hfil     &
 \hfil Empty space  \hfil \\&& \\
 Evolution&\hfil  Governed by the field equations \hfil &  \hfil  ? \hfil\\
 &\hspace*{0.4\textwidth}\\
 Grav.\ field &   Determined by (field eqs + Einstein eqs)  &  \hfil
 Rigid  \hfil
 \\
\cline{2-3}\rule{0pt}{15pt}%
Origin  & \multicolumn{2}{c}{Topological obstructions} \\
&\multicolumn{2}{c}{}
\\
 Cosmological r\^ole& \multicolumn{2}{c}{Wakes}\\ &\multicolumn{2}{c}{}\\
 Manifestations
  & \multicolumn{2}{c}{Multiple images of a single source}
\end{tabular}
 \end{center}
\end{table}
In contrast to the strings, the singularities are in the \emph{empty}
space. So, their evolution is
\emph{not} connected to properties of any field and, in particular, is
not described by the Nambu action. The cosmic strings, unlike the
singularities, can bend, curving the spacetime around (in particular,
they can emit gravitational waves). On the other hand, both entities have
similar mechanisms of formation and stability, both produce wakes and,
finally both may give rise to multiple images of a single source.

To appreciate the latter property
\begin{figure}[h,t,b]
\psfrag{E}{$\eucl^3$}
\psfrag{p}{$p$}
\begin{center}
\includegraphics[width=0.7\textwidth]{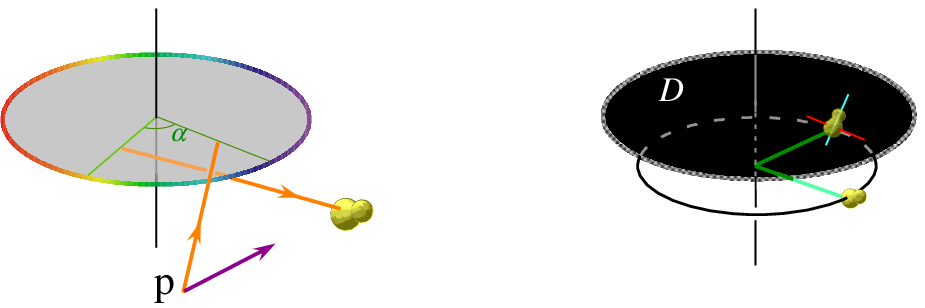}
\end{center}
\hspace*{9em} (a)\hfill (b)\hspace*{9em}
\caption{\label{fig:lensing} (a) The line of sight (the orange line) of
the observer located in $p$ ends up at the object which he also sees when
looks in the direction shown by the purple arrow. (b) The object seen
through the disk looks rotated (w.~r.~t.\ all three axes --- blue, green,
and red) in comparison with the same object observed  directly.}
\end{figure}
consider the loop singularity from Example~\ref{ex} and an observer who
looks  in the direction of the singularity, see Fig~\ref{fig:lensing}a.
His line of sight after reaching the disk jumps, changes its direction
and may end on an object which he could see somewhere aside. Thus he will
see \emph{two} images of the same source
--- one  in the direction shown by the orange ray and another in the
direction of the purple ray. Equivalently, one can pick a source seen out
of the disk and rotate it in one's mind by $\alpha$, see
Fig~\ref{fig:lensing}b. If it gets behind the disk one will see both
images. Note that generally the images differ, because one sees the
object from different angles.

Thus, suppose there is a loop string-like singularity somewhere. And
suppose  there is a galaxy not far from it,
\begin{figure}[h,b,t]
\psfrag{a}{(a)}
\psfrag{b}{(b)}
\psfrag{c}{(c)}
\psfrag{d}{(d)}
\begin{center}
\includegraphics[width=0.8\textwidth]{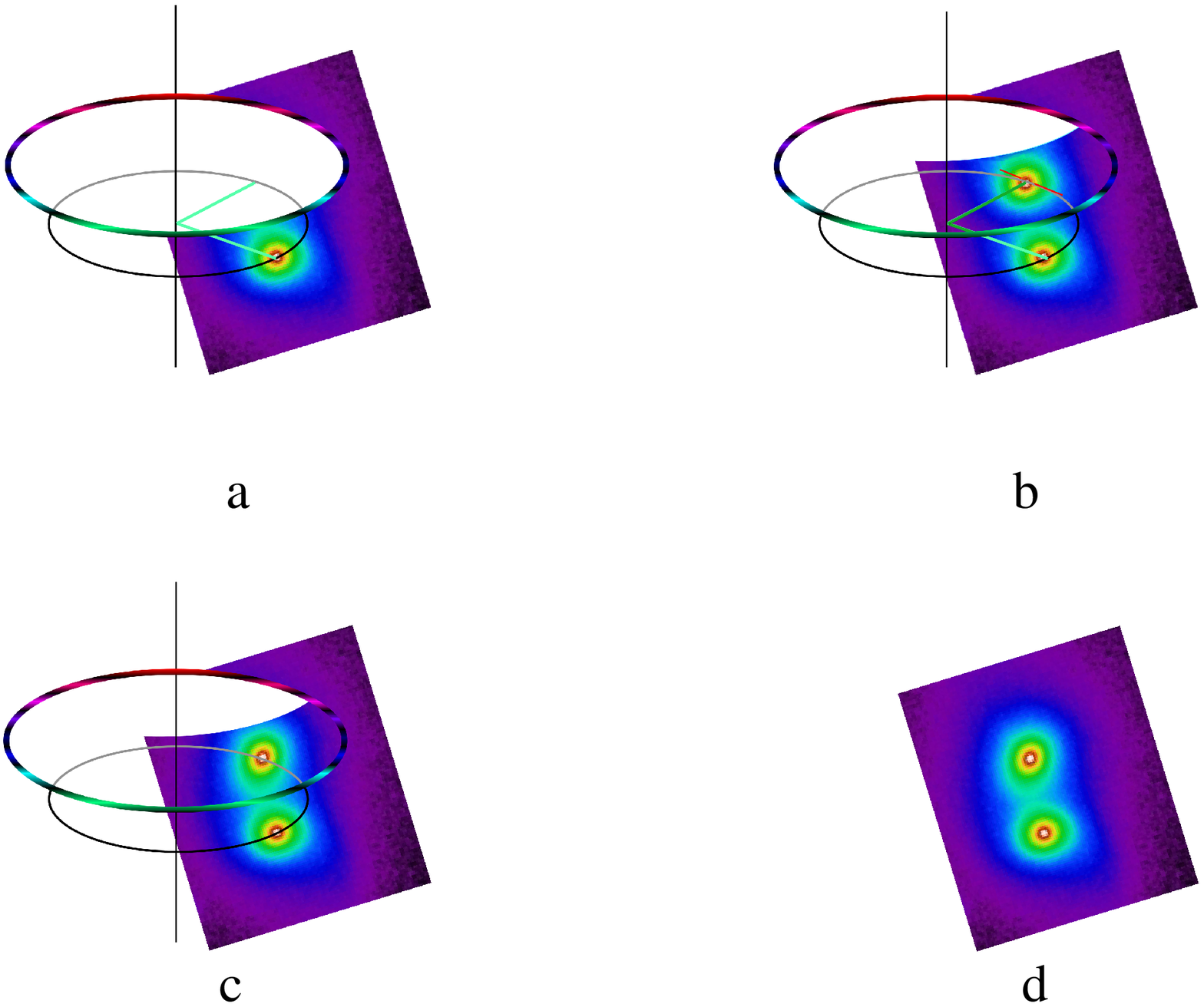}
\end{center}
\caption{\label{fig:observ} (c) is what we must see if there is a galaxy
near a loop string-like singularity as in (a). (d) The image of the
extragalactic object CSL-1.}
\end{figure}
see Fig.~\ref{fig:observ}a.  Then following the prescription given above
we rotate the galaxy by $\alpha$ w.~r.~t.\ the axis of the disk $D$
corresponding to the singularity, and find that (for a suitable size of
the loop and  value of $\alpha$) a terrestrial observer would see two
images of the galaxy, see Fig.~\ref{fig:observ}b. The second image must
be rotated w,~r.~t. to the other one to make allowance for the effect
mentioned above, so, it will look as in Fig.~\ref{fig:observ}c. Thus, if
there are galactic size loop string-like singularities, one may expect to
observe in the sky something like that depicted in
Fig.~\ref{fig:observ}c. And Fig.~\ref{fig:observ}d is the
\emph{real image} of the extragalactic object CSL-1 obtained with the
Hubble Space Telescope, see \cite{Sazh} and references therein.
Apparently the object is a pair of giant galaxies with the same
velocities, with the same (at the 98\% c.l.) spectra and with no
explanation\footnote{As is argued in \cite{Sazh} this cannot be a result
of lensing by a compact object (because the isophotes are undistorted) or
by a straight string (because no other pairs are found nearby).} for this
similarity other than sheer coincidence.

\end{document}